\documentstyle[prd,aps,preprint,tighten,epsfig]{revtex}

\begin{document}

\draft

\title{Possible Effects of Noncommutative Geometry \\
on Weak $CP$ Violation and Unitarity Triangles}
\author{{\bf Zhe Chang} ~ and ~ {\bf Zhi-zhong Xing}}
\address{CCAST (World Laboratory), P.O. Box 8730, Beijing 100080, China \\
and Institute of High Energy Physics, Chinese Academy of Sciences, \\
P.O. Box 918 (4), Beijing 100039, China 
\footnote{Mailing address} \\
({\it Electronic address: changz@mail.ihep.ac.cn; xingzz@mail.ihep.ac.cn}) }
\maketitle

\begin{abstract}
Possible effects of noncommutative geometry on weak $CP$ violation and
unitarity triangles are discussed by taking account of a simple
version of the momentum-dependent quark mixing matrix in the 
noncommutative standard model. In particular, we calculate nine
rephasing invariants of $CP$ violation and illustrate the noncommutative 
$CP$-violating effect in a couple of charged $D$-meson decays. We also
show how inner angles of the {\it deformed} unitarity triangles are related
to $CP$-violating asymmetries in some typical $B_d$ and $B_s$ transitions
into $CP$ eigenstates. $B$-meson factories are expected to help probe
or constrain noncommutative geometry at low energies in the near future.
\end{abstract}


\section{Introduction}

One of the major goals of $B$-meson factories is to test the 
Kobayashi-Maskawa mechanism of $CP$ violation in the 
standard model (SM) \cite{KM}. 
If this mechanism is correct, all $CP$-violating asymmetries in weak
decays of quark flavors must be proportional to a universal
and rephasing-invariant parameter $\cal J$ \cite{Jarlskog}, defined through
\begin{eqnarray}
{\cal J}^{ij}_{\alpha\beta} & \equiv & {\rm Im} 
\left ( V_{\alpha i} V_{\beta j} V^*_{\alpha j} V^*_{\beta i} \right )  
\nonumber \\
& = & {\cal J} \sum_{\gamma, k} \left ( \epsilon_{\alpha\beta\gamma} 
\epsilon_{ijk} \right ) \; ,
%	(1.1)
\end{eqnarray}
where $V$ denotes the Cabibbo-Kobayashi-Maskawa (CKM) matrix of
quark flavor mixing, and its Greek and Latin subscripts 
run respectively over $(u, c, t)$ and $(d, s, b)$. A number of
promising measurables of $CP$ violation at $B$-meson factories
are directly related to the unitarity triangle shown in Fig. 1(a), 
which describes the following orthogonal relation of $V$ in the 
complex plane:
\begin{equation}
V^*_{ub} V_{ud} + V^*_{cb} V_{cd} + V^*_{tb} V_{td} = 0 \; .
%	(1.2)
\end{equation}
The inner angles of this unitarity triangle are commonly defined as
\begin{eqnarray}
\alpha & \equiv & \arg \left (- \frac{V^*_{tb}V_{td}}{V^*_{ub}V_{ud}}
\right ) \; ,
\nonumber \\
\beta & \equiv & \arg \left (- \frac{V^*_{cb}V_{cd}}{V^*_{tb}V_{td}}
\right ) \; ,
\nonumber \\
\gamma & \equiv & \arg \left (- \frac{V^*_{ub}V_{ud}}{V^*_{cb}V_{cd}}
\right ) \; .
%	(1.3)
\end{eqnarray}
Of course, $\alpha + \beta + \gamma = \pi$ and 
${\cal J} \propto \sin\alpha \propto \sin\beta \propto \sin\gamma$ hold.
So far the $CP$-violating asymmetry in $B^0_d$ vs $\bar{B}^0_d
\rightarrow J/\psi K_{\rm S}$ decays, which approximates to $\sin 2\beta$
to a high degree of accuracy in the SM, has been unambiguously measured
at both KEK and SLAC \cite{Beta}. 
Further experiments are expected to help determine 
all three angles of the unitarity triangle and test the consistency of
the Kobayashi-Maskawa picture of $CP$ violation.

Another major goal of $B$-meson factories is to detect possible new
sources of $CP$ violation beyond the SM. On the one hand,
the Kobayashi-Maskawa mechanism of $CP$ violation is unable to generate
a sufficiently large matter-antimatter asymmetry of the universe observed 
today; and on the other hand, many extensions of the SM do allow the 
presence of new $CP$-violating phenomena \cite{New}. Therefore it is
well-motivated to look for new sources of $CP$ violation in various 
weak decays of quark (and lepton) flavors. A particularly interesting
possibility is that new $CP$ violation may stem from noncommutative geometry.

Noncommutative geometry plays a very important role in unraveling 
properties of the Planck-scale physics.
It has for a long time been suspected that the noncommutative spacetime 
might be a realistic picture of how
spacetime behaves near the Planck scale \cite{05}. 
Strong quantum fluctuations of gravity may make points fuzzy. In fact,
the noncommutative geometry naturally enters the theory of open string 
in a background $B$-field \cite{06}.
In particular, the noncommutative geometry makes the holography \cite{09} 
(e.g., the AdS/CFT correspondence) of a higher dimensional quantum 
system of gravity and
lower dimensional theory possible. It was also discovered that simple limits
of $M$ theory and superstring theory lead directly to the 
noncommutative gauge field theory \cite{07,08}.  The fluctuations of the
$D$-brane are described by the noncommutative gauge field theory \cite{10}.  
The noncommutative field theory has been intensively
studied in the past two decades \cite{11}. A standard model on noncommutative 
spacetime was even set up \cite{12}. However, in recent years,
the study of noncommutative geometry has been focused on the so-called 
Moyal plane, with the coordinates and their conjugate
momenta satisfying the relations \cite{13}
\begin{eqnarray}
\left [x^\mu \stackrel{\star}{,} x^\nu \right ] & = & i\theta^{\mu\nu} ~, 
\nonumber \\
\left [x^\mu \stackrel{\star}{,} p^\nu \right ] & = & i\hbar\eta^{\mu\nu} ~,
%	(1.4)
\end{eqnarray}
where  $\theta^{\mu\nu}$ is a constant antisymmetric matrix. 
Here the Moyal-Weyl star product can be defined by a
formal power series,
\begin{equation}
(f\star g)(x) \; = \; e^{\frac{i}{2}\theta^{\mu\nu}
\frac{\partial}{\partial x^\mu}\frac{\partial}{\partial x^\nu}}
f(x)g(y)\vert_{x=y} ~.
%	(1.5)
\end{equation}
There are two obstacles in the way of building a SM-like 
gauge field theory on the Moyal plane. The first one
is the charge quantization in the noncommutative QED \cite{14}. 
The charges of matter fields coupled to the $U_\star (1)$ gauge
theory are fixed to only three possible
values, $\pm 1$ and $0$, depending on the representation of particles.
This is indeed a problem in view of the range of hypercharges in 
the $U(1)_Y$ part of the SM. The second one is due to  
extra $U_\star (1)$ gauge 
fields \cite{15}. Under the infinitesimal gauge transformation
$\hat{\delta}$, the vector gauge potential $\hat{V}_\mu$, 
the fundamental matter field $\hat{\Psi}$ and the Higgs
field $\hat{\Phi}$ transform as
\begin{eqnarray}
\hat{\delta}\hat{V}_\mu & = & \partial_\mu\hat{\Lambda} +
i[\hat{\Lambda}\stackrel{\star}{,}\hat{V}_\mu] ~,
\nonumber \\
\hat{\delta}\hat{\Psi} & = & i\hat{\Lambda} \star \hat{\Psi} ~,
\nonumber \\
\hat{\delta}\hat{\Phi} & = & i\hat{\Lambda} \star \hat{\Phi} - i\hat{\Phi}
\star \hat{\Lambda}' ~.
%	(1.6)
\end{eqnarray}
It should be noticed that the Moyal-Weyl product would destroy the 
closure condition of the $SU_\star (n)$. For example, two
Lie algebra-valued consecutive transformations 
$\hat{\delta}_{\hat{\Lambda}}(=\Lambda_a(x)T^a)$ and
$\hat{\delta}_{\hat{\Lambda}'}
(=\Lambda'_aT^a)$ of the matter fields in
the fundamental representation,
\begin{equation}
[\hat{\delta}_{\hat{\Lambda}} \stackrel{\star}{,}
\hat{\delta}_{\hat{\Lambda}'}] \; =\; 
\frac{1}{2}\{\Lambda_a(x)\stackrel{\star}{,}
\Lambda'_b(x)\}[T^a,T^b] + \frac{1}{2}[\Lambda_a(x)
\stackrel{\star}{,}\Lambda'_b(x)]\{T^a,T^b\} ~,
%	(1.7)
\end{equation}
are not equivalent to a Lie algebra-valued gauge transformation. 
The only group which admits a simple noncommutative
extension is $U(N)$. However, there are
extra $U_\star (1)$ factors in the $U_\star (N)$ gauge field theory 
compared to the extended SM on the noncommutative space. 
In order to construct an $SU_\star (3)\times SU_\star (2)\times U_\star (1)$ 
Yang-Mills theory \cite{16}, Wess and his collaborators 
\cite{17,18,19,20} have extended the ordinary 
Lie algebra-valued gauge transformations to
enveloping algebra-valued noncommutative
gauge transformations,
\begin{equation}
\hat{\Lambda} \; =\; \Lambda_a^0(x)T^a + \Lambda_{ab}^1:T^aT^b: +
\Lambda^2_{abc}(x):T^aT^bT^c: + \cdots ~,
%	(1.8)
\end{equation}
where $:T^{a_1}T^{a_2}\cdots T^{a_m}:$ denotes a symmetric ordering 
under the exchange of the index $a_i$. This kind of
extension of the gauge transformations and the Seiberg-Witten 
map \cite{06} together solves the two main problems in
building a noncommutative SM quite well.

The purpose of this paper is to examine possible effects of
noncommutative geometry on weak $CP$ violation and CKM unitarity
triangles. In section II, we elucidate a simple version of the 
momentum-dependent CKM matrix in the noncommutative SM, which consists of
a new source of $CP$ violation induced by nonvanishing $\theta^{\mu\nu}$.
We calculate the rephasing invariants of $CP$ violation in section III,
and find that the noncommutative $CP$-violating effects may be 
manifest in a couple of charged $D$-meson decays. In section IV,
we show how the CKM unitarity triangles in the SM get modified 
in the noncommutative SM. We also figure out the relations between 
inner angles of the {\it deformed} unitarity triangles and $CP$-violating 
asymmetries in some nonleptonic decays of $B_d$ and $B_s$ mesons.
Section V is devoted to a brief summary of our main results.

\section{Momentum-dependent CKM matrix}
\setcounter{equation}{0}

The noncommutative SM \cite{21,22} is an 
$SU_\star (3)\times SU_\star (2)\times U_\star (1)$
gauge field theory on the Moyal plane,
\small
\begin{equation}
S_{\rm YM} = -\int d^4x\left[\frac{1}{2g'} 
{\rm Tr}_{u(1)}(\hat{F}_{\mu\nu} \star \hat{F}^{\mu\nu}) +
\frac{1}{2g} {\rm Tr}_{su(2)}(\hat{F}_{\mu\nu} \star \hat{F}^{\mu\nu}) +
\frac{1}{2g^{~}_S} {\rm Tr}_{su(3)}(\hat{F}_{\mu\nu} 
\star \hat{F}^{\mu\nu})\right] ~.
%	(2.1)
\end{equation}
\normalsize
The gauge field strength $\hat{F}_{\mu\nu}$ is given by
\begin{equation}
\hat{F}_{\mu\nu} = \partial_\mu \hat{V}_\nu - \partial_\nu 
\hat{V}_\mu - i[\hat{V}_\mu\stackrel{\star}{,}\hat{V}_\nu] ~,
%	(2.2)
\end{equation}
where $\hat{V}_\mu$ is the vector potential of the 
$SU_\star (3)\times SU_\star (2)\times U_\star (1)$
gauge field, which is related to the ordinary potential
\begin{equation}
V_\mu = g'A_\mu(x)Y + g\sum_{a=1}^3 B_{\mu a}(x)T_L^a +
g^{~}_S\sum_{a=1}^8 G_{\mu a}(x)T^a_S ~,
%	(2.3)
\end{equation}
by the Seiberg-Witten map (to the first order of $\theta^{\mu\nu}$)
\begin{equation}
\hat{V}_\mu = V_\mu + \frac{1}{4}\theta^{\lambda\rho}
\{V_\rho,\partial_\lambda V_\mu\} + \frac{1}{4}\theta^{\lambda\rho}
\{F_{\lambda\mu},V_\nu\} + {\cal O}(\theta^2) ~.
%	(2.4)
\end{equation}
Here $F^{\mu\nu} = \partial^\mu V^\nu-\partial^\nu V^\mu - i[V^\mu,V^\nu]$
is the ordinary field strength,
and $Y$, $T_L^a$ and $T_S^a$ are the generators of $U(1)_Y$, $SU(2)_L$ and 
$SU(3)_C$ respectively.

The parameter $\hat{\Lambda}$ of the gauge transformations on the 
noncommutative space is determined by the ordinary
gauge parameter $\Lambda$ via the Seiberg-Witten map,
\begin{equation}
\hat{\Lambda} = \Lambda + \frac{1}{4}\theta^{\mu\nu}
\{V_\nu,\partial_\mu \Lambda\} + {\cal O}(\theta^2) ~,
%	(2.5)
\end{equation}
where the ordinary gauge parameter $\Lambda$ is of the form
\begin{equation}
\Lambda = g'\tau(x)Y + g\sum_{a=1}^3 \tau_a^L(x)T_L +
g^{~}_S\sum_{a=1}^8\tau_a^S(x)T_S^a ~.
%	(2.6)
\end{equation}
The Seiberg-Witten maps for the Higgs field $\hat{\Phi}$
and the fermion field $\hat{\Psi}$ are given as
\begin{eqnarray}
\hat{\Phi} & = & \Phi+\frac{1}{2}\theta^{\mu\nu}V_\nu
\left [\partial_\mu\Phi-\frac{i}{2}(V_\mu\Phi-\Phi V_\mu') \right ]
+ \frac{1}{2}\theta^{\mu\nu} \left [\partial_\mu\Phi - \frac{i}{2}
(V_\mu\Phi - \Phi V_\mu') \right ] V'_\nu + {\cal O}(\theta^2) ~,
\nonumber \\
\hat{\Psi} & = & \Psi + \frac{1}{2}\theta^{\mu\nu}V_\nu\partial_\mu\Psi +
\frac{i}{8}\theta^{\mu\nu}[V_\mu,V_\nu]\Psi
+ {\cal O}(\theta^2) ~.
%	(2.7)
\end{eqnarray}
At this stage, we can say that a SM-like gauge field 
theory on the noncommutative spacetime is set up
consistently. Many interesting properties of noncommutative 
spacetime can be investigated directly within the
framework of the noncommutative SM \cite{22,others}.

In the noncommutative SM, the $W$-quark-quark $SU(2)_L$ vertex 
in the flavor basis can be written as
\begin{equation}
{\cal L}_{Wqq} \; = \; \overline{(u' ~~~ c' ~~~ t')_L} ~ J_{\rm cc} 
\left (\matrix{
d' \cr s' \cr b' \cr} \right )_L + ~ {\rm h.c.} ~,
%	(2.8)
\end{equation}
where the superscript ``prime'' denotes the flavor or interaction eigenstates
of quarks, and 
\begin{equation}
J_{\rm cc} = \frac{\sqrt{2}}{2}g\gamma^\mu W^+_\mu - ig\frac{\sqrt{2}}{4}
\left(\frac{1}{2}\theta^{\mu\nu}\gamma^\alpha + \theta^{\nu\alpha}
\gamma^\mu\right)
(\partial_\mu W_\nu^+ - \partial_\nu W^+_\mu)\partial_\alpha
%	(2.9)
\end{equation}
represents the charged current. Note that the charged-current interactions 
with more than one $W^\pm$ and (or) $Z$ bosons as well as those
with gluons \cite{21} are not included in Eqs. (II.8) and (II.9), 
since they are not
closely associated with our subsequent discussions about weak $CP$ violation 
and unitarity triangles. To diagonalize the Yukawa interactions of quarks 
with the Higgs boson, one should make proper unitary rotations on the up- 
and down-type quark fields. In the basis where the Yukawa coupling matrices 
are diagonal, the $W$-quark-quark $SU(2)_L$ vertex in Eq. (II.8) becomes
\begin{equation}
{\cal L}_{Wqq} \; = \; \overline{(u ~~~ c ~~~ t)_L} ~ U J_{\rm cc} 
\left (\matrix{
d \cr s \cr b \cr} \right )_L + ~ {\rm h.c.} ~.
%	(2.10)
\end{equation}
Within the SM (i.e., $\theta^{\mu\nu} = 0$), $U$ turns out to be the
CKM matrix $V$ after a spontaneous breakdown of the $SU(2)_L$ symmetry.

Making use of the antisymmetric property of $\theta^{\mu\nu}$ and 
taking account of the $SU(2)_L$ symmetry, we have the following
relations for the $W$-$u$-$d$ vertex:
\begin{eqnarray}
\int d^4x \left [ \overline{u(p)}\theta^{\nu\alpha}\gamma^\mu 
\partial_\mu W_\nu^+\partial_\alpha d(q) \right ]
& = &-\int d^4x \left [ \overline{u(p)}\theta^{\nu\alpha}\gamma^\mu
(p_\mu-q_\mu) q_\alpha W_\nu^+ d(q) \right ] 
\nonumber \\
& = & 0 ~,
%	(2.11)
\end{eqnarray}
and
\begin{eqnarray}
\int d^4x \left [ \overline{u(p)}\frac{1}{2}\theta^{\mu\nu}\gamma^\alpha
(\partial_\mu W_\nu^+ - \partial_\nu W_\mu^+)
\partial_\alpha d(q) \right ]
& = &-\int d^4x \left [ \overline{u(p)}\theta^{\mu\nu}(p_\mu-q_\mu)
\gamma^\alpha q_\alpha W^+_\nu d(q) \right ] 
\nonumber \\
& = & 0 ~.
%	(2.12)
\end{eqnarray}
Therefore, we can generally rewrite the $W$-quark-quark $SU(2)$ vertex 
in the form
\begin{equation}
{\cal L}_{Wqq} \; = \; \frac{\sqrt{2}}{2}g 
~ \overline{(u ~~~ c ~~~ t)_L} ~ \overline{U} \gamma^\mu W^+_\mu
\left (\matrix{
d \cr s \cr b \cr} \right )_L + ~ {\rm h.c.} ~,
%	(2.13)
\end{equation}
where we have used the notation
\begin{equation}
\overline{U}_{\alpha k}(p, q) \; = \; 
U_{\alpha k} \left (1 ~ - ~ \frac{i}{2} ~ p^\mu_\alpha 
\theta_{\mu\nu} q^\nu_k \right ) ~,
%	(2.14)
\end{equation}
with $\alpha$ and $k$ running respectively over $(u, c, t)$ and $(d, s, b)$. 
The momentum-dependent
matrix $\overline{U}$ is not guaranteed to be unitary, and its
new phases (induced by non-zero $\theta^{\mu\nu}$) may lead to
new $CP$-violating effects in weak interactions.
 
Indeed the afore-mentioned property of $\overline{U}$ has been observed 
by Hinchliffe and Kersting in Ref. \cite{Hinchliffe}. They point out
that the signal for noncommutative geometry at low energies can simply
be a momentum-dependent CKM matrix $\overline{V}$, which is defined
in analogy with $\overline{U}$ as follows:
\begin{equation}
\overline{V} \; = \; V -
\frac{i}{2} \left ( \matrix{
V_{ud} x_{ud}	
& V_{us} x_{us}	
& V_{ub} x_{ub} \cr
V_{cd} x_{cd}	
& V_{cs} x_{cs}	
& V_{cb} x_{cb} \cr
V_{td} x_{td}	
& V_{ts} x_{ts}	
& V_{tb} x_{tb} \cr}
\right ) \; ,
%	(2.15)
\end{equation}
where $x_{\alpha k} \equiv p^\mu_\alpha \theta_{\mu \nu} q^\nu_k$ 
for $\alpha = u, c, t$ and $k = d, s, b$. This {\it effective}
flavor mixing matrix arises from an approximation of the exact
noncommutative SM in the leading order of $\theta^{\mu\nu}$. 
Subsequently we explore some phenomenological implications of 
$\overline{V}$ on weak $CP$ violation and unitarity triangles. 

\section{Rephasing invariants of $CP$ violation}
\setcounter{equation}{0}

The momentum-dependent CKM matrix $\overline{V}$ is not 
unitary in general, as one can see from Eq. (II.15).
Note that the following normalization relations hold up to 
${\cal O}(x^2_{\alpha i})$:
\begin{eqnarray}
\sum_\alpha |\overline{V}_{\alpha i}|^2 & = & 
\sum_\alpha |V_{\alpha i}|^2 =  1 \; ,
\nonumber \\
\sum_i |\overline{V}_{\alpha i}|^2 & = & 
\sum_i |V_{\alpha i}|^2 = 1 \; .
%	(3.1)
\end{eqnarray}
On the other hand, we obtain ($i\neq j$ and $\alpha \neq \beta$)
\begin{eqnarray}
\sum_\alpha (\overline{V}^*_{\alpha i} \overline{V}_{\alpha j}) & = & 
i \sum_\alpha \left [ \left (V^*_{\alpha i} V_{\alpha j} \right )
\frac{x_{\alpha i} -x_{\alpha j}}{2} \right ] ,
\nonumber \\
\sum_i (\overline{V}^*_{\alpha i} \overline{V}_{\beta i}) & = & 
i \sum_i \left [ \left (V^*_{\alpha i} V_{\beta i} \right ) 
\frac{x_{\alpha i} -x_{\beta i}}{2} \right ] ,
%	(3.2)
\end{eqnarray}
which do not vanish unless $(x_{\alpha i} -x_{\alpha j}) = {\rm constant}$
and $(x_{\alpha i} -x_{\beta i}) = {\rm constant}$. 

The observables of $CP$ violation in the noncommutative SM
must depend upon the imaginary parts of nine rephasing invariants
$(\overline{V}_{\alpha i} \overline{V}_{\beta j}
\overline{V}^*_{\alpha j} \overline{V}^*_{\beta i})$. Up to 
${\cal O}(x_{\alpha i})$, we have
\begin{eqnarray}
\overline{\cal J}^{ij}_{\alpha\beta} & \equiv & 
{\rm Im} \left ( \overline{V}_{\alpha i} \overline{V}_{\beta j} 
\overline{V}^*_{\alpha j} \overline{V}^*_{\beta i} \right )  
\nonumber \\
& = & {\cal J} 
\sum_{\gamma, k} \left ( \epsilon_{\alpha\beta\gamma} \epsilon_{ijk} \right ) 
~ + ~ {\cal R}^{ij}_{\alpha\beta} \xi^{ij}_{\alpha\beta} \; , 
%	(3.3)
\end{eqnarray}
where 
\begin{eqnarray}
{\cal R}^{ij}_{\alpha\beta} & \equiv & {\rm Re} \left (V_{\alpha i} 
V_{\beta j} V^*_{\alpha j} V^*_{\beta i} \right ) \; ,
\nonumber \\
\xi^{ij}_{\alpha\beta} & \equiv & \frac{1}{2} \left ( x_{\alpha j}
+ x_{\beta i} - x_{\alpha i} - x_{\beta j} \right ) \; ;
%	(3.4)
\end{eqnarray}
and the subscripts $(\alpha, \beta, \gamma)$ and $(i, j, k)$ run respectively 
over $(u,c,t)$ and $(d,s,b)$. If $\overline{V}$ were unitary (i.e., 
$\xi^{ij}_{\alpha\beta} = 0$), the term associated with 
${\cal R}^{ij}_{\alpha \beta}$ would vanish and the equality 
$\overline{\cal J}^{ij}_{\alpha\beta} = {\cal J}^{ij}_{\alpha\beta}$ 
would hold. Otherwise, both the magnitude and the sign of
$\overline{\cal J}^{ij}_{\alpha\beta}$ rely on the momentum-dependent parameter
$\xi^{ij}_{\alpha\beta}$ which signifies the effect of noncommutative 
geometry. To get an order-of-magnitude feeling about the SM and 
noncommutative SM contributions to $\overline{\cal J}^{ij}_{\alpha\beta}$, 
we adopt the Wolfenstein
parametrization \cite{Wolfenstein} for the CKM matrix $V$ and then obtain
\begin{equation}
{\cal J} \; \approx \; A^2 \lambda^6 \eta \; ,
%	(3.5)
\end{equation}
and
\begin{eqnarray}
{\cal R}^{ds}_{uc} & \approx & - \lambda^2 \; ,
\nonumber \\
{\cal R}^{ds}_{ut} & \approx & - A^2 \lambda^6 \left (1 - \rho \right ) \; ,
\nonumber \\
{\cal R}^{ds}_{ct} & \approx & A^2 \lambda^6 \left (1 - \rho \right ) \; ,
\nonumber \\
{\cal R}^{db}_{uc} & \approx & - A^2 \lambda^6 \rho \; ,
\nonumber \\
{\cal R}^{db}_{ut} & \approx & A^2 \lambda^6 \left [ \rho 
\left (1 - \rho \right ) - \eta^2 \right ] \; ,
\nonumber \\
{\cal R}^{db}_{ct} & \approx & - A^2 \lambda^6 \left (1 - \rho \right ) \; ,
\nonumber \\
{\cal R}^{sb}_{uc} & \approx & A^2 \lambda^6 \rho \; ,
\nonumber \\
{\cal R}^{sb}_{ut} & \approx & - A^2 \lambda^6 \rho \; ,
\nonumber \\
{\cal R}^{sb}_{ct} & \approx & - A^2 \lambda^4 \; ,
%	(3.6)
\end{eqnarray}
where $A \approx 0.81$, $\lambda \approx 0.22$, $\rho \approx 0.15$ and
$\eta \approx 0.34$ extracted from a global fit of current experimental 
data in the framework of the SM \cite{PDG}. The values of $\lambda$
and $A$, which are extracted respectively from the semileptonic
$K_{e3}$ and $B\rightarrow \bar{D}^{(*)}l^+\nu^{~}_l$ decays without
any loop-induced pollution, keep unchanged even in the presence of
noncommutative geometry. In contrast, $\rho$ and $\eta$ are sensitive to
possible new physics induced by loop (box and penguin) effects, which
may reside in $B^0_d$-$\bar{B}^0_d$ mixing, $B^0_s$-$\bar{B}^0_s$ mixing 
and $CP$ violation in $K^0$-$\bar{K}^0$ mixing. 
Hence the results of $\rho$ and $\eta$ are expected to deviate somehow
from their SM values in a new analysis of current experimental data, 
when the noncommutative SM takes the place of the SM. It is unnecessary
to know an accurate range of $\rho$ or $\eta$, however, for our purpose 
to illustrate the effects of noncommutative geometry on weak $CP$ violation
and unitarity triangles.
One can see that ${\cal J} \ll |{\cal R}^{sb}_{ct}| 
\ll |{\cal R}^{ds}_{uc}|$ holds, while the other seven
${\cal R}^{ij}_{\alpha\beta}$ have comparable sizes as ${\cal J}$.
Note in particular that
\begin{eqnarray}
\overline{\cal J}^{ds}_{uc} & \approx & A^2 \lambda^6 \eta - \lambda^2
\xi^{ds}_{uc} \; ,
\nonumber \\
\overline{\cal J}^{sb}_{ct} & \approx & A^2 \lambda^6 \eta - A^2 \lambda^4
\xi^{sb}_{ct} \; .
%	(3.7)
\end{eqnarray}
Thus the noncommutative $CP$-violating effect may be comparable with or 
dominant over the SM one, if $\xi^{ds}_{uc}$ is of ${\cal O}(\lambda^4)$
or larger in $\overline{\cal J}^{ds}_{uc}$; 
and if $\xi^{sb}_{ct}$ is of ${\cal O}(\lambda^2)$ or larger
in $\overline{\cal J}^{sb}_{ct}$.

To see how the rephasing invariants $\overline{\cal J}^{ij}_{\alpha\beta}$ 
are related to $CP$-violating asymmetries in specific weak decays, 
let us take $D^\pm_s \rightarrow K^\pm K_{\rm S}$ for example. Direct
$CP$ violation arises from the interference between the Cabibbo-allowed
channel and the doubly Cabibbo-suppressed channel of $D^\pm_s$ decays into
the final states $K^\pm K_{\rm S}$, where $K^0$-$\bar{K}^0$ mixing leads
to an additional $CP$-violating effect of magnitude
$2{\rm Re}\epsilon^{~}_K \approx 3.3\times 10^{-3}$ \cite{Xing95}. 
The latter dominates over the former in the SM, because 
two interferring amplitudes of $D^+_s$ or $D^-_s$ transitions have a small
relative weak phase 
$\arg [ (V_{cd}V^*_{ud})/(V_{cs}V^*_{us})] \approx A^2\lambda^4\eta
\sim 5 \times 10^{-4}$
and a small relative size $|V_{cd}V^*_{us}|/|V_{cs}V^*_{ud}| \approx
\lambda^2 \sim 5\times 10^{-2}$ \cite{Lipkin}:
\begin{eqnarray}
A(D^+_s \rightarrow K^+ K_{\rm S}) & \propto &
\left ( V_{cs} V^*_{ud} \right ) q^*_K + 
\left ( V_{cd} V^*_{us} \right ) p^*_K R_s ~ e^{{\rm i} \delta_s} \; ,
\nonumber \\
A(D^-_s \rightarrow K^- K_{\rm S}) & \propto &
\left ( V^*_{cs} V_{ud} \right ) p^*_K + 
\left ( V^*_{cd} V_{us} \right ) q^*_K R_s ~ e^{{\rm i} \delta_s} \; ,
%	(3.8)
\end{eqnarray}
where $p^{~}_K$ and $q^{~}_K$ are the $K^0$-$\bar{K}^0$ mixing parameters
%%%%%%%%%%%%%%%%%
\footnote{Since $CP$ violation in the kaon system is tiny, 
we expect that the weak phase of $K^0$-$\bar{K}^0$ mixing is nearly the 
same as that of $K^0$ vs $\bar{K}^0$ decays, which amounts to 
$(V_{us}V^*_{ud})/(V^*_{us}V_{ud})$ at the tree level \cite{Cohen}. 
It is therefore plausible to take $q^{~}_K/p^{~}_K = 
[(V_{us}V^*_{ud})(1-\epsilon^{~}_K)]/[(V^*_{us}V_{ud})(1+\epsilon^{~}_K)]$ 
as an effective description of the weak phase and the associated
$CP$ violation in $K^0$-$\bar{K}^0$ mixing.},
%%%%%%%%%%%%%%%%%%
$\delta_s$ denotes the relative strong phase difference
between two interferring decay amplitudes, and 
$R_s \approx 1 + a_2/a_1 \approx -1.2$ in the factorization
approximation for relevant hadronic matrix elements
($a_1 \approx 1.1$ and $a_2 \approx -0.5$ being the effective Wilson 
coefficients at the ${\cal O}(m_c)$ scale \cite{BSW}). 
When noncommutative geometry is taken into consideration, the relative
weak phase between two interferring decay amplitudes of $D^+_s$ or
$D^-_s$ meson becomes associated with 
${\rm Im} [(\overline{V}_{cd}\overline{V}^*_{ud})/
(\overline{V}_{cs}\overline{V}^*_{us})]$. In this case,
we obtain the momentum-dependent
$CP$-violating asymmetry between the partial rates of 
$D^-_s \rightarrow K^-K_{\rm S}$ and
$D^+_s \rightarrow K^+K_{\rm S}$ decays as follows:
\begin{eqnarray}
{\cal A}_s & \equiv & \frac{|A (D^-_s \rightarrow K^-K_{\rm S})|^2
- |A (D^+_s \rightarrow K^+K_{\rm S})|^2}
{|A (D^-_s \rightarrow K^-K_{\rm S})|^2 +
|A (D^+_s \rightarrow K^+K_{\rm S})|^2} \; 
\nonumber \\
& \approx & 2 {\rm Re}\epsilon^{~}_K - 2 \overline{\cal J}^{ds}_{uc}
R_s \sin\delta_s \; .
%	(3.9)
\end{eqnarray}
If $\delta_s \sim {\cal O}(1)$ and $\xi^{ds}_{uc} \sim {\cal O}(\lambda^2)$ 
or $\overline{\cal J}^{ds}_{uc} \sim {\cal O}(\lambda^4)$ held,
two different contributions to ${\cal A}_s$ would be comparable in
magnitude. Therefore a significant deviation of ${\cal A}_s$ from 
$2{\rm Re}\epsilon^{~}_K$, if experimentally observed, would signal the 
presence of new physics, which is likely to be noncommutative geometry.

\section{Unitarity triangles in $B$-meson decays}
\setcounter{equation}{0}

In the complex plane,
the vector $\overline{V}^*_{\alpha i} \overline{V}_{\beta i}$ can be 
obtained from rotating the vector $V^*_{\alpha i} V_{\beta i}$ 
anticlockwise to a small angle $(x_{\alpha i} -x_{\beta i})/2$. It is
therefore expected that 
$\overline{V}^*_{ub} \overline{V}_{ud}$, 
$\overline{V}^*_{cb} \overline{V}_{cd}$ and 
$\overline{V}^*_{tb} \overline{V}_{td}$ do not form a close triangle,
as shown in Fig. 1(b).
Nevertheless, one may define three angles by using these three vectors:
\begin{eqnarray}
\overline{\alpha} & \equiv & 
\arg \left (- \frac{\overline{V}^*_{tb} \overline{V}_{td}}
{\overline{V}^*_{ub} \overline{V}_{ud}} \right ) \; ,
\nonumber \\
\overline{\beta} & \equiv & \arg \left (- \frac{\overline{V}^*_{cb}
\overline{V}_{cd}}{\overline{V}^*_{tb} \overline{V}_{td}}
\right ) \; ,
\nonumber \\
\overline{\gamma} & \equiv & \arg \left (- \frac{\overline{V}^*_{ub}
\overline{V}_{ud}}{\overline{V}^*_{cb} \overline{V}_{cd}}
\right ) \; .
%	(4.1)
\end{eqnarray} 
Comparing between Eqs. (I.3) and (IV.1), we find
\begin{eqnarray}
\overline{\alpha} & = & \alpha + \xi^{db}_{tu} \; ,
\nonumber \\
\overline{\beta} & = & \beta + \xi^{db}_{ct} \; ,
\nonumber \\
\overline{\gamma} & = & \gamma + \xi^{db}_{uc} \; .
%	(4.2)
\end{eqnarray}
By definition in Eq. (III.4), 
$\xi^{db}_{tu} + \xi^{db}_{ct} + \xi^{db}_{uc} = 0$ holds.
It turns out that
\begin{equation}
\overline{\alpha} + \overline{\beta} + \overline{\gamma} 
= \alpha + \beta + \gamma = \pi
%	(4.3)
\end{equation}
holds too. In Ref. \cite{Hinchliffe}, the momentum-dependent features
of $\overline{\alpha}$, $\overline{\beta}$ and $\overline{\gamma}$
are illustrated in the assumption of $\eta =0$ or ${\cal J} =0$
(i.e., $CP$ violation from the SM is switched off).

Besides $\alpha$, $\beta$ and $\gamma$, $CP$ violation in weak $B$-meson 
decays is also associated with the following three angles of the
CKM unitarity triangles in the SM \cite{Cohen}:
\begin{eqnarray}
\gamma' & \equiv & \arg\left (-\frac{V^*_{ub}V_{tb}}{V^*_{us}V_{ts}}
\right ) \; ,
\nonumber \\
\delta & \equiv & \arg\left (-\frac{V^*_{tb}V_{ts}}{V^*_{cb}V_{cs}}
\right ) \; ,
\nonumber \\
\omega & \equiv & \arg\left (-\frac{V^*_{us}V_{ud}}{V^*_{cs}V_{cd}}
\right ) \; .
%	(4.4)
\end{eqnarray}
It is easy to check that the relation
$\delta + \omega = \gamma - \gamma'$ holds. The counterparts of 
$\gamma'$, $\delta$ and $\omega$ in the noncommutative SM are
defined as 
\begin{eqnarray}
\overline{\gamma}' & \equiv & \arg\left (-\frac{\overline{V}^*_{ub}
\overline{V}_{tb}}{\overline{V}^*_{us}\overline{V}_{ts}}\right ) \; ,
\nonumber \\
\overline{\delta} & \equiv & \arg\left (-\frac{\overline{V}^*_{tb}
\overline{V}_{ts}}{\overline{V}^*_{cb}\overline{V}_{cs}}\right ) \; ,
\nonumber \\
\overline{\omega} & \equiv & \arg\left (-\frac{\overline{V}^*_{us}
\overline{V}_{ud}}{\overline{V}^*_{cs}\overline{V}_{cd}}\right ) \; .
%	(4.5)
\end{eqnarray}
Of course, the similar relation $\overline{\delta} + \overline{\omega} 
= \overline{\gamma} - \overline{\gamma}'$ holds.
Comparing between Eqs. (IV.4) and (IV.5), we obtain
\begin{eqnarray}
\overline{\gamma}' & = & \gamma' + \xi^{sb}_{ut} \; ,
\nonumber \\
\overline{\delta} & = & \delta + \xi^{sb}_{tc} \; ,
\nonumber \\
\overline{\omega} & = & \omega + \xi^{ds}_{uc} \; .
%	(4.6)
\end{eqnarray}
One can see that $\omega$ or $\overline{\omega}$ is actually
the weak phase associated with $D^\pm_s \rightarrow K^\pm K_{\rm S}$
decays discussed above. 
As $|\delta| \approx \lambda^2\eta \sim 2\times 10^{-2}$ and
$|\omega| \approx A^2\lambda^4\eta \sim 5\times 10^{-4}$ in the SM,
the noncommutative effect is possible to be comparable with 
$\delta$ in $\overline{\delta}$ and dominant over $\omega$ in 
$\overline{\omega}$. In particular, the latter could be a sensitive window
to probe or constrain noncommutative geometry at low energies.

The weak angles $\overline{\alpha}$, $\overline{\beta}$, $\overline{\gamma}$,
$\overline{\gamma}'$, $\overline{\delta}$ and $\overline{\omega}$ can
be determined from direct and indirect $CP$-violating asymmetries in 
a variety of weak $B$ decays. Here let us consider neutral 
$B_d$ and $B_s$ decays into $CP$ eigenstates.
In the neglect of penguin-induced pollution, indirect 
$CP$ violation in such decay modes may arise from the interplay of direct 
$B^0_q$ and $\bar{B}^0_q$ decays (for $q= d$ or $s$) and
$B^0_q$-$\bar{B}^0_q$ mixing \cite{Du}. If the final state consists of 
$K_{\rm S}$ or $K_{\rm L}$ meson, then $K^0$-$\bar{K}^0$ mixing should also
be taken into account. In the box-diagram approximation of the SM,
the weak phase of $B^0_q$-$\bar{B}^0_q$ mixing is associated with 
the CKM factor $(V^*_{tb}V_{tq})/(V_{tb}V^*_{tq})$. On the other hand,
the weak phase of $K^0$-$\bar{K}^0$ mixing can simply be taken as
$(V_{us}V^*_{ud})/(V^*_{us}V_{ud})$, since $CP$ violation is tiny
in the kaon system \cite{Cohen}.
When noncommutative geometry is concerned, all $V_{\alpha i}$ should
be replaced by $\overline{V}_{\alpha i}$. 

To illustrate how the inner angles of {\it deformed} unitarity triangles are 
related to the $CP$-violating asymmetries in neutral $B$-meson decay
modes, we take $B^0_d$ vs $\bar{B}^0_d \rightarrow J/\psi K_{\rm S}$ and
$B^0_s$ vs $\bar{B}^0_s \rightarrow J/\psi K_{\rm S}$ transitions for
example. Their indirect $CP$-violating asymmetries $\Delta_d$ and
$\Delta_s$ are given respectively as 
\begin{eqnarray}
\Delta_d & = & - {\rm Im} \left (\frac{\overline{V}^*_{tb}\overline{V}_{td}}
{\overline{V}_{tb}\overline{V}^*_{td}} \cdot \frac{\overline{V}_{cb}
\overline{V}^*_{cs}}{\overline{V}^*_{cb}\overline{V}_{cs}}
\cdot \frac{\overline{V}_{us}\overline{V}^*_{ud}}
{\overline{V}^*_{us}\overline{V}_{ud}} \right ) \; = \; 
+ \sin 2 (\overline{\beta} + \overline{\omega}) \; , 
\nonumber \\
\Delta_s & = & - {\rm Im} \left (\frac{\overline{V}^*_{tb}\overline{V}_{ts}}
{\overline{V}_{tb}\overline{V}^*_{ts}} \cdot \frac{\overline{V}_{cb}
\overline{V}^*_{cd}}{\overline{V}^*_{cb}\overline{V}_{cd}}
\cdot \frac{\overline{V}^*_{us}\overline{V}_{ud}}
{\overline{V}_{us}\overline{V}^*_{ud}} \right ) \; = \; 
- \sin 2 (\overline{\gamma} - \overline{\gamma}') \; . 
%	(4.7)
\end{eqnarray}
Here we have taken into account the fact that 
$J/\psi K_{\rm S}$ is a $CP$-odd state. 
Possible deviations of such momentum-dependent observables from the
SM predictions are worth searching for at $B$-meson factories.

In Table 1, we list a number of typical decay channels of $B_d$ and
$B_s$ mesons and their $CP$-violating asymmetries, including two
examples given above. One can see that the weak angles 
$\overline{\alpha}$, $\overline{\beta}$, $\overline{\gamma}$,
$\overline{\gamma}'$, $\overline{\delta}$ and $\overline{\omega}$
are (in principle) measurable. The self-consistent relations
such as $\overline{\alpha} + \overline{\beta} + \overline{\gamma} = \pi$
and $\overline{\gamma} - \overline{\gamma}' = \overline{\delta} +
\overline{\omega}$ could be tested, if the relevant angles were able to
be determined at the same momentum scale. 

Note that it is possible to distinguish noncommutative geometry 
from some other sources of new physics in indirect $CP$-violating
asymmetries of $B_d$ and $B_s$ decays. Taking a variety of supersymmetric
standard models for example, we find that those
$CP$-violating asymmetries listed in Table 1 may get corrections 
from gauginos, Higgsinos and squarks through box diagrams which produce
nonstandard $\Delta B =2$ effects. This kind of new physics can be
parametrized in terms of two phases \cite{Cohen},
\begin{eqnarray}
\theta_d & \equiv & \frac{1}{2} \arg \left (
\frac{\langle B^0_d |{\cal H}^{\rm full}_{\rm eff}|\bar{B}^0_d\rangle}
{\langle B^0_d |{\cal H}^{\rm SM}_{\rm eff}|\bar{B}^0_d\rangle} 
\right ) \; ,
\nonumber \\
\theta_s & \equiv & \frac{1}{2} \arg \left (
\frac{\langle B^0_s |{\cal H}^{\rm full}_{\rm eff}|\bar{B}^0_s\rangle}
{\langle B^0_s |{\cal H}^{\rm SM}_{\rm eff}|\bar{B}^0_s\rangle} 
\right ) \; ,
%	(4.8)
\end{eqnarray}
where ${\cal H}^{\rm full}_{\rm eff}$ is the effective Hamiltonian
consisting of both standard and supersymmetric contributions, and
${\cal H}^{\rm SM}_{\rm eff}$ consists only of the contribution from the
SM box diagrams. In the presence of $\theta_d$ and $\theta_s$,
the $CP$-violating asymmetries given in Table 1 get modified.
We list the new results for those asymmetries in Table 2. 
Comparing between Tables 1 and 2, one can see that noncommutative
and supersymmetric effects are actually distinguishable, if some
of those $CP$-violating asymmetries are measured at $B$-meson
factories. In particular,
the $CP$ asymmetry in $B^0_s$ vs $\bar{B}^0_s \rightarrow \eta'\eta'$ 
decay modes is nonvanishing only if there is new physics contributing
to $B^0_s$-$\bar{B}^0_s$ mixing.

In our discussions, the penguin-induced effects have been neglected.
This approximation is expected to be reasonable for those $B^0_d$ and
$B^0_s$ transitions
occurring through the subprocesses $\bar{b}\rightarrow \bar{c} c \bar{d}$ 
and $\bar{b}\rightarrow \bar{c} c \bar{s}$, but it might be 
problematic for those charmless decay modes whose tree-level
amplitudes are strongly CKM-suppressed. In the latter case, significant 
new noncommutative $CP$-violating effects may appear via the
QCD penguins as a result of $CP$-odd phase factors in the relevant 
quark-gluon vertices \cite{21}. The entanglement of different types 
of noncommutative $CP$ violation in weak decays of
quark flavors should be carefully analyzed. Such an analysis, which
must involve much complexity and subtlety of the noncommutative SM, is
beyond the scope of this paper.

\section{Summary}
\setcounter{equation}{0}

We have examined possible effects of noncommutative geometry on weak
$CP$ violation and unitarity triangles based on a simple version of
the momentum-dependent CKM matrix in the noncommutative SM. 
Among nine rephasing invariants of $CP$ violation, we find that two of them
are sensitive to the noncommutative corrections. In particular, the
noncommutative $CP$-violating effect could be comparable with or 
dominant over the SM one in $D^\pm_s \rightarrow K^\pm K_{\rm S}$ decays.
We have also illustrated how the CKM unitarity triangles get deformed 
in the noncommutative SM. Simple relations are established between 
inner angles of the {\it deformed} unitarity triangles and $CP$-violating 
asymmetries in some typical decays of $B_d$ and $B_s$ mesons into $CP$
eigenstates, such as $B_d \rightarrow J/\psi K_{\rm S}$ and
$B_s \rightarrow D^+_s D^-_s$. We anticipate that $B$-meson factories may
help probe or constrain noncommutative geometry at low energies in the
near future.

Finally we remark that further progress in the noncommutative 
gauge field theory will allow us to study the phenomenology of 
noncommutative geometry on a more solid ground. 

\acknowledgments{
We like to thank X. Calmet and X.G. He for useful
discussions. This work was supported in part by the National
Natural Science Foundation of China.}

\newpage

%%%%%%%%%%%%%%%%%%%% Fig. 1 %%%%%%%%%%%%%%%%
\begin{figure}
\vspace{-1cm}
\epsfig{file=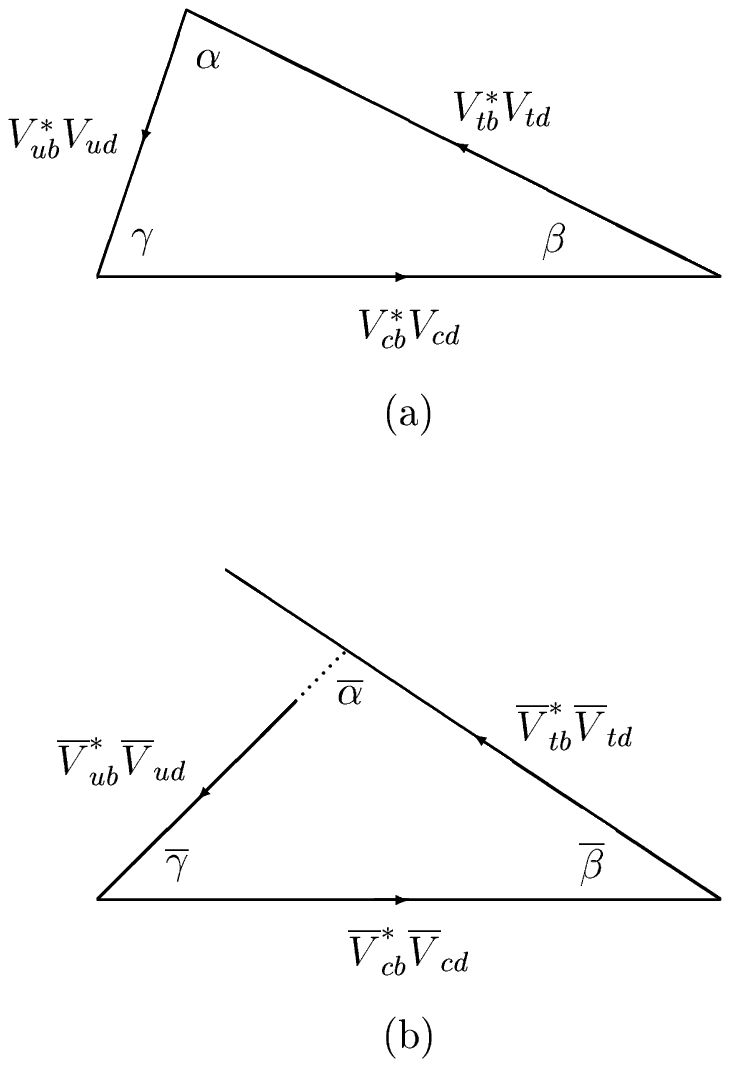,bbllx=1cm,bblly=4cm,bburx=20cm,bbury=32cm,%
width=14cm,height=22cm,angle=0,clip=}
\vspace{-9.6cm}
\caption{The CKM unitarity triangle in the standard model (a) 
and its {\it deformed} counterpart in the noncommutative standard model (b).}
\end{figure}
%%%%%%%%%%%%%%%%%%%%%%%%%%%%%%%%%%%%%%%%%%%%

\newpage

%%%%%%%%%%%%%%%%%% Table 1 %%%%%%%%%%%%%%%%%%%%
\begin{table}[t]
\caption{Typical $B_d$ and $B_s$ decays and associated
$CP$-violating asymmetries in the noncommutative standard model.}
\begin{center}
\begin{tabular}{lllllll} \\
Class &~~~& Sub-process &~~~& Decay mode &~~~& $CP$ asymmetry \\ \\ \hline \\ 
%-----------------------------------
1d 
&& $\bar{b} \rightarrow \bar{c}c\bar{s}$
&&
$B^0_d\rightarrow J/\psi K_{\rm S}$ 
&& $+\sin 2 (\overline{\beta} + \overline{\omega})$
\\ \\
%------------------------------------------
2d 
&& $\bar{b} \rightarrow \bar{c}c\bar{d}$
&&
$B^0_d\rightarrow D^+D^-$ 
&& $-\sin 2 \overline{\beta}$
\\ \\
%------------------------------------------
3d 
&& $\bar{b} \rightarrow \bar{u}u\bar{d}$
&&
$B^0_d\rightarrow \pi^+\pi^-$ 
&& $+\sin 2 \overline{\alpha}$
\\ \\
%------------------------------------------
4d 
&& $\bar{b} \rightarrow \bar{s}s\bar{s}$
&&
$B^0_d\rightarrow \phi K_{\rm S}$ 
&& $-\sin 2 (\overline{\alpha} + \overline{\gamma}')$
\\ \\ \hline \\
%=================================================
1s
&& $\bar{b} \rightarrow \bar{c}c\bar{s}$
&&
$B^0_s\rightarrow D^+_s D^-_s$ 
&& $+\sin 2 \overline{\delta}$
\\ \\
%------------------------------------------
2s
&& $\bar{b} \rightarrow \bar{c}c\bar{d}$
&&
$B^0_s\rightarrow J/\psi K_{\rm S}$ 
&& $-\sin 2 (\overline{\gamma} - \overline{\gamma}')$
\\ \\
%------------------------------------------
3s
&& $\bar{b} \rightarrow \bar{u}u\bar{d}$
&&
$B^0_s\rightarrow \rho K_{\rm S}$ 
&& $+\sin 2 \overline{\gamma}'$
\\ \\
%------------------------------------------
4s
&& $\bar{b} \rightarrow \bar{s}s\bar{s}$
&&
$B^0_s\rightarrow \eta' \eta'$ 
&& $0$
\\ \\ 
%-----------------------------------------------------------
\end{tabular}
\end{center}
\end{table}
%%%%%%%%%%%%%%%%%%%%%%%%%%%%%%%%%%%%%%%%%%%%%%%%%%%%%%%%%%%%%%%%%%%%%%%%%%%%

%%%%%%%%%%%%%%%%%% Table 2 %%%%%%%%%%%%%%%%%%%%
\begin{table}[t]
\caption{Typical $B_d$ and $B_s$ decays and associated
$CP$-violating asymmetries in the noncommutative standard model
and in the presence of supersymmetric $\Delta B =2$ effects.}
\begin{center}
\begin{tabular}{lllllll} \\
Class &~~~& Sub-process &~~~& Decay mode &~~~& $CP$ asymmetry \\ \\ \hline \\ 
%-----------------------------------
1d 
&& $\bar{b} \rightarrow \bar{c}c\bar{s}$
&&
$B^0_d\rightarrow J/\psi K_{\rm S}$ 
&& $+\sin 2 (\overline{\beta} + \overline{\omega} + \theta_d)$
\\ \\
%------------------------------------------
2d 
&& $\bar{b} \rightarrow \bar{c}c\bar{d}$
&&
$B^0_d\rightarrow D^+D^-$ 
&& $-\sin 2 (\overline{\beta} + \theta_d)$
\\ \\
%------------------------------------------
3d 
&& $\bar{b} \rightarrow \bar{u}u\bar{d}$
&&
$B^0_d\rightarrow \pi^+\pi^-$ 
&& $+\sin 2 (\overline{\alpha} - \theta_d)$
\\ \\
%------------------------------------------
4d 
&& $\bar{b} \rightarrow \bar{s}s\bar{s}$
&&
$B^0_d\rightarrow \phi K_{\rm S}$ 
&& $-\sin 2 (\overline{\alpha} + \overline{\gamma}' - \theta_d)$
\\ \\ \hline \\
%=================================================
1s
&& $\bar{b} \rightarrow \bar{c}c\bar{s}$
&&
$B^0_s\rightarrow D^+_s D^-_s$ 
&& $+\sin 2 (\overline{\delta} - \theta_s)$
\\ \\
%------------------------------------------
2s
&& $\bar{b} \rightarrow \bar{c}c\bar{d}$
&&
$B^0_s\rightarrow J/\psi K_{\rm S}$ 
&& $-\sin 2 (\overline{\gamma} - \overline{\gamma}' + \theta_s)$
\\ \\
%------------------------------------------
3s
&& $\bar{b} \rightarrow \bar{u}u\bar{d}$
&&
$B^0_s\rightarrow \rho K_{\rm S}$ 
&& $+\sin 2 (\overline{\gamma}' + \theta_s)$
\\ \\
%------------------------------------------
4s
&& $\bar{b} \rightarrow \bar{s}s\bar{s}$
&&
$B^0_s\rightarrow \eta' \eta'$ 
&& $-\sin 2\theta_s$
\\ \\ 
%-----------------------------------------------------------
\end{tabular}
\end{center}
\end{table}
%%%%%%%%%%%%%%%%%%%%%%%%%%%%%%%%%%%%%%%%%%%%%%%%%%%%%%%%%%%%%%%%%%%%%%%%%%%%

\end{document}